\RequirePackage[l2tabu, orthodox]{nag} 
\RequirePackage{fix-cm} 

\documentclass[dvipsnames,svgnames]{article}

\usepackage[ruled,linesnumbered]{algorithm2e}

\SetAlFnt{\small}
\SetAlCapFnt{\small}
\SetAlCapNameFnt{\small}
\SetAlCapHSkip{0pt}
\IncMargin{-\parindent}

\usepackage[hyphens]{url}
\usepackage{hyperref}
\hypersetup{breaklinks=true}

\usepackage[usenames,dvipsnames,svgnames,table]{xcolor}

\usepackage{graphicx}

\usepackage{complexity}

\usepackage{amsmath} 
\usepackage{amssymb} 
\usepackage{amsfonts} 
\usepackage{amsthm} 

\usepackage[a4paper]{geometry}

\usepackage{mathrsfs}

\usepackage{etoolbox}

\usepackage{microtype}

\usepackage{booktabs}

\usepackage[]{nicefrac}

\usepackage{nth}

\usepackage{paralist}

\usepackage{xspace}

\usepackage{listings}
\usepackage{rotating}

\usepackage{caption}
\usepackage{subcaption}
\usepackage[section]{placeins}

\usepackage{authblk}

\usepackage{pgfplots}
\pgfplotsset{compat=1.12}
\usepgfplotslibrary{units}
\usepackage{forest}
\definecolor{folderbg}{RGB}{124,166,198}
\definecolor{folderborder}{RGB}{110,144,169}
\def\Size{4pt}
\tikzset{
	folder/.pic={
		\filldraw[draw=folderborder,top color=folderbg!50,bottom color=folderbg]	(-1.05*\Size,0.2\Size+5pt) rectangle ++(.75*\Size,-0.2\Size-5pt);
		\filldraw[draw=folderborder,top color=folderbg!50,bottom color=folderbg]
		(-1.15*\Size,-\Size) rectangle (1.15*\Size,\Size);
		\filldraw[draw=folderborder,top color=folderbg!50,bottom color=folderbg]
		(-1.15*\Size,-\Size) rectangle (1.15*\Size,\Size);
	}
}
\usetikzlibrary{fit}
\usetikzlibrary{positioning}

\usepackage{tikz-uml}
\usetikzlibrary{positioning}
\usetikzlibrary{spy}

\usepackage{booktabs}

\newbool{use_todo}
\setbool{use_todo}{true}
\ifbool{use_todo}{
	\usepackage[colorinlistoftodos, shadow, textsize=small, textwidth=35mm]{todonotes}
	\setlength{\marginparwidth}{3.5cm}
	}{
	\usepackage[disable]{todonotes}
}

\newcommand{\ts}[1]{\texttt{\scriptsize{#1}}}

\renewcommand{\A}{\mathcal{A}}
\newcommand{\I}{\mathcal{I}}
\renewcommand{\C}{\mathcal{C}}

\renewcommand{\S}{\mathcal{S}}

\newcommand{\init}{\texttt{initialize}}
\newcommand{\topp}{\texttt{top}}
\newcommand{\eof}{\texttt{EOF}}
\newcommand{\readin}{\texttt{readInput}}

\newcommand{\popc}{\texttt{popCondition}}
\newcommand{\prepop}{\texttt{prePop}}
\newcommand{\pop}{\texttt{pop}}
\newcommand{\postpop}{\texttt{postPop}}
\newcommand{\nopop}{\texttt{noPop}}

\newcommand{\pushc}{\texttt{pushCondition}}
\newcommand{\prepush}{\texttt{prePush}}
\newcommand{\push}{\texttt{push}}
\newcommand{\postpush}{\texttt{postPush}}
\newcommand{\nopush}{\texttt{noPush}}

\newcommand{\report}{\texttt{report}}

\providecommand{\keywords}[1]{\textbf{\textit{Keywords---}} #1}

\usepackage[capitalise]{cleveref}
\crefname{subsection}{subsection}{subsections}

\newtheorem{problem}{Problem}
\crefname{problem}{problem}{problems}

\usepackage[comma,square,sort&compress,numbers]{natbib}
\renewcommand{\harvardurl}[1]{\textbf{URL:} \url{#1}} 

\begin{document}

\title{Experimental Study of Compressed Stack Algorithms in Limited Memory Environments}
\author[1,2]{Jean-Fran\c{c}ois Baffier\thanks{J.-F.~B.~was supported by JST ERATO Grant Number JPMJER1305, Japan}}
\author[3]{Yago Diez\thanks{Y.~D.~was supported by the IMPACT Tough Robotics Challenge Project of Japan Science and Technology Agency}}
\author[3]{Matias Korman\thanks{M.~K.~was supported in part by the ELC project (MEXT KAKENHI No.~12H00855, 15H02665, and 17K12635)}}
\affil[1]{National Institute of Informatics}
\affil[2]{JST--ERATO Kawarabayashi Large Graph project}
\affil[3]{Tohoku University}


\maketitle

\begin{abstract}
 The {\em compressed stack} is a data structure designed by Barba {\em et al.} (Algorithmica 2015) that allows to reduce the amount of memory needed by an algorithm (at the cost of increasing its runtime). In this paper we introduce the first  implementation of this data structure and make its source code publicly available.

 Together with the implementation we analyze the performance of the compressed stack. In our synthetic experiments, considering different test scenarios and using data sizes ranging up to $2^{30}$ elements, we compare it with the classic (uncompressed) stack, both in terms of runtime and memory used.

 Our experiments show that the compressed stack needs significantly less memory than the usual stack (this difference is significant for inputs containing $2000$ or more elements). Overall, with a proper choice of parameters, we can save a significant amount of space (from two to four orders of magnitude) with a small increase in the runtime ($2.32$ times slower on average than the classic stack). These results holds even in test scenarios specifically designed to be challenging for the compressed stack.\newline

 \noindent\keywords{Stack algorithms, time-space trade-off, convex hull, implementation}
\end{abstract}

\section{Introduction}\label{sec:intro}
In recent years we have seen a huge growth in the everyday use of small devices such as smartphones, sensor networks, or even security cameras. These devices are becoming more and more powerful, but due to several constraints (ranging from budget issues to discouraging possible theft) it is sometimes desirable to keep them small in both memory size and computational power.

Consequently, theoretical computer science is expressing a renewed interest in the design of algorithms that use little space. A recent trend in the community has been the appearance of {\em time-space trade-off} algorithms~\citep{abbkmrs-mcasp-11}; that is, algorithms that can take into account space constraints; the larger the amount of memory that they have available, the less time that the algorithms will need. We refer the interested reader to~\citep{k-mca-15} for a survey on the different models that have been proposed to handle space constraints.

From a theoretical point of view the interest has been in the relationship between time and space. In most cases, the dependency has been linear or almost linear~\citep{kmrrss-tstotvd-17,akprr-tstotsp-17,kmrrss-tstotvd-15,abbkmrs-mcasp-12,bkls-cvpufv-13}: that is, when we double the amount of available space we expect the runtime to more or less halve.

We believe that all these theoretical contributions are reaching a point where they can be used in practice. As a result, in this paper we take a more hands-on approach on the topic. Rather than studying theoretical dependences, we implement one of the proposed approaches to time-space trade-off and thoroughly assess its behaviour when executed using benchmark data of varying difficulty. Specifically, we are interested on seeing how much we can reduce the amount of memory consumed by algorithms while we make sure that runtimes remain reasonable.

Among the several results in this field, we focus on the \emph{compressed stack} data structure introduced by \citet{barba2014spacetime}. The main reason, among several others, to choose this data structure for our analysis is that it is useful for several algorithms (as opposed to most techniques that are only be used for a specific problem); this data structure can be used to reduce the amount of space used by any deterministic incremental algorithm whose internal structure is a stack (see more details in \Cref{sec:preliminaries}).

Another good property of this structure is that, once properly implemented, the algorithm is unaware of which data structure it utilizes: it suffices to replace the classic (uncompressed) stack data structure with a compressed stack. This modular transparency makes it easy for users to adopt, and ideal for comparison purposes. Also, the dependency between time and space is not the same everywhere: for small amounts of memory the dependency is exponential (that is, by increasing the memory by a small constant we can {\em halve} the runtime), but it quickly becomes logarithmic afterwards (we need to {\em double} the amount of space to see any difference in the runtimes).

Thus, our study has two objectives. First, we want to verify that the theoretical dependency between time and space actually matches practice. Also, we want to provide some guidelines on how to find this breakpoint in the dependency so that the user can choose the right amount of memory to achieve the faster algorithm that fits their memory constraints.

\subsection{Results and Paper Organization}
Our main contribution is the implementation of the compressed stack data structure of \citet{barba2014spacetime}. The implementation is freely available at \citep{baffierCPPcompressed}. With the use of this library, one can implement any algorithm that uses this data structure quickly and efficiently (see examples of succinctness of the structure in \cref{subsec:examples}).

In \cref{sec:preliminaries} we give a brief overview of stack algorithms and the compressed stack data structure. In \Cref{sec:implementation} we give a quick overview of how to use our library and discuss some minor differences between our implementation and the theoretical formulation by Barba {\em et al.}.

In \Cref{sec:xp} we present a thorough study on the behaviour of the compressed stack in favourable and unfavourable scenarios (both constructed with synthetic data). We pay special attention to the comparison between the (theoretical) expected behaviour and the actual results obtained.

As expected, the compressed stack structure uses significantly less memory than the classic stack. From the theory we know that the running times must increase, but only a rough idea on the increase can be told. From the experiments done in this paper, we can deduce guidelines for prospective users so that the amount of memory is drastically reduced while we keep the runtimes relatively low. Further discussion about parameter settings is done in \Cref{sec:conclusion}.


\section{Preliminaries}\label{sec:preliminaries}

The compressed stack data structure can only be used with a family of algorithms (called {\em stack algorithms}). This class  includes widely used algorithms addressing problems such as computing the convex hull of a set of points, approximating a histogram by a unimodal function, or computing the visibility region of a point inside a polygonal domain. See~\cite{barba2014spacetime,banerjee2015timespace} for more examples of stack algorithms.

In full generality, we look at algorithms whose the input is a list of elements $\I=\{a_1, \ldots, a_n\}$, and the goal is to find a subset of $\I$ that satisfies a previously defined property. In a nutshell, we are looking at deterministic incremental algorithms that use a stack, and possibly other small data structures $\C$ (this additional structure is called the \emph{context} and ideally only consists in a few integers).

A stack algorithm solves the problem in an incremental fashion, scanning the elements of $\I$ one by one. At any point during the execution, the stack keeps the values that form the solution up to that point. For each new element $a$ that is taken from $\I$, the algorithm pops all values of the stack that do not satisfy a predefined "pop condition" and if $a$ meets some other "push condition", it is pushed into the stack. The algorithm then proceeds to the next element in  $\I$ until all elements  have been processed. The final result is normally contained in the stack, and at the end it is reported. This is done by simply reporting the top of the stack, popping the top vertex, and repeating until the stack is empty. Thus, an algorithm $\A$ that follows this scheme is called a \emph{stack} algorithm (see a pseudo-code in \cref{algo:theory}).
\begin{algorithm}[!htp]
	\caption{Theoretical scheme of a stack algorithm}\label{algo:theory}

	Initialize stack $\S$ and context $\C$ (auxiliary data structure) with $O(1)$ elements from $\I$\\
	\ForAll{subsequent input $a\in \I$\nllabel{line:theoryread}}{
		\While{$\A$.\popc($a$,$\C$,$\A$.\topp($1$),\ldots,$\A$.\topp($k$))\nllabel{line:theorystartpop}}{
			$\A$.\pop()\nllabel{line:theorypop}
			}\nllabel{line:theoryendpop}
		\If{$\A$.\pushc($a$,$\C$,$\A$.\topp($1$),\ldots,$\A$.\topp($k$))\nllabel{line:theorystartpush}}{
			$\A$.\push(a)\nllabel{line:theorypush}
			}\nllabel{line:theoryendpush}
	}
	$\A$.\report()
\end{algorithm}

\subsection{Sample problem: convex hull computation}\label{subsec:hull_descri}
A typical example of a stack algorithm is the convex hull problem: given a list of points $p_1,\ldots, p_n$ in the plane sorted in increasing values of their $x$-coordinate, we want to compute their convex hull, i.e., the smallest convex set that encloses all of the them. Among the many algorithms that solve this problem, ~\citep{a-hltchasp}\footnote{The algorithms in this survey actually compute a slightly more general problem: computing the convex hull of a simple polygon, but both problems are almost identical. The simple polygon case has a few more difficult cases (such as when the polygon spirals around itself), but they have no impact on the way in which the stack is handled. Thus, for simplicity we only describe the simpler case.}, the one by~\citep{l-ofchsp-83} falls in the class of stack algorithms.

\begin{figure}[t]
 \centering
 \includegraphics[width=0.7\textwidth]{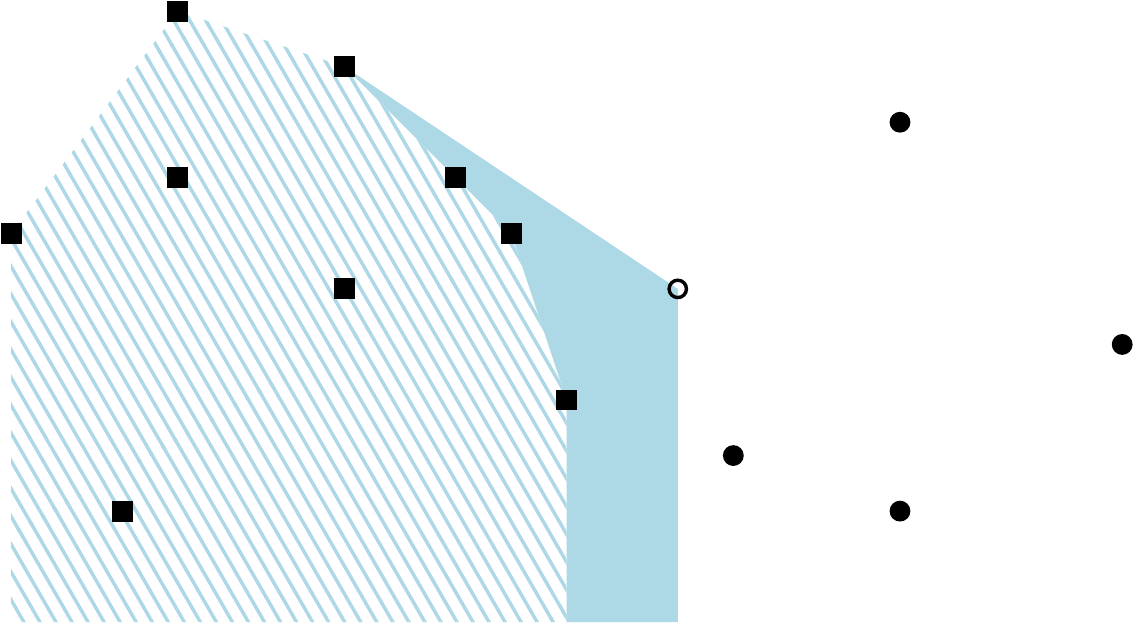}
 \caption{One step of Lee's algorithm: the upper convex hull after processing 9 leftmost points (denoted by squares in the figure) and their convex hull is highlighted (dashed region). The next point (empty circle) is a witness that three points do not belong to the hull. These four points were the last ones to be added into the upper hull and thus they are popped from the stack. The updated hull is also shown (in solid).}\label{fig:convex}
\end{figure}

Lee's algorithm processes the points sequentially and stores the elements that are currently candidates for being in the convex hull. For simplicity, we discuss how to compute the upper hull (i.e., points that are in the convex hull and are above the line passing through the rightmost and leftmost point) of a set of points\footnote{The traditional algorithm scans once the input to compute the upper hull  and a second time to compute the lower hull, but we note that the same algorithm can be modified to work in one pass by using two stacks. In any case, neither of these two options have a large impact on the overall working of the stack algorithm, so we ignore this and focus in the upper hull only.}.

When a new element is processed it may be witness to several points that were previously in the upper hull and should not be there anymore (see \cref{fig:convex}). The key property is that those points must be the last ones that were considered as candidates. Consequently, they are removed in reverse order of insertion and thus a stack is the perfect data structure to store the list of candidates.

We refer the interested reader to~\citep{bkos-cgaa-00} for more details on the convex hull problem and its applications. As an illustration on the simplicity in implementing stack algorithms, we have implemented this algorithm as part of our library (details are given in \cref{subsec:examples}).

\subsection{Compressed Stack Overview}\label{subsec:cpoverview}

In this section we briefly describe the compressed stack data structure making a special emphasis on how can we save memory. During the execution of a stack algorithm, one may have many elements of the input in the stack. These elements are stored explicitly if we use the traditional stack resulting in high memory usage. 

In the compressed stack structure, the user chooses a parameter $p$ (to indicate the amount of space that the algorithm is allowed to use). Then the input is split into $p$ blocks. Whenever a block has been fully processed (i.e., the incremental algorithm has scanned the last element of the block) we {\em compress} the block: rather than explicitly storing it, we store a small amount of information\footnote{there exist two ways in which the block can be compressed, totally and partially. The basic concept is introduced in \cref{subsec:csimpl}. We refer the reader to \citet{barba2014spacetime} for complete details on this}.

This saves a lot of memory, since a block could have many elements in the stack but only a fixed amount of information per block is stored instead. Since we scan the input in a monotone fashion only one block can be partially processed at any instant of time. We store that block and its preceding block {\em explicitly} (i.e., most of the information is explicitly stored), while all other ones are somehow compressed.

Stack algorithms only need access to the top of the stack at any given time. Since this information is known by the compressed stack, we can perform as a usual stack. Eventually, the algorithm may pop many elements, and then the information inside a block that was compressed will be needed. This information can be \emph{reconstructed} by re-executing the same algorithm, but only restricted to a portion of the input. The key trick to keeping the runtime small is to make sure that few reconstructions are needed, and always restricted to small portions of the input.

We emphasize that the working of the compressed stack is {\em transparent} to the algorithm. The algorithm is running, does some push and pop operations as well as reading the top of the stack. The stack data structure handles compression of information independently. Sometime during the execution, a pop will trigger a reconstruction. In this moment, the algorithm is paused and we launch a copy of the same algorithm (with a smaller input). Once the small execution ends, the needed information is available in memory, and we can resume with the main execution.

From a theory standpoint, stack algorithms run in $O(n^{1+\frac{1}{\log p}})$ time using $O(p\log_p n)$ space for any parameter $p\in\lbrace 2, \ldots n\}$. In particular, when $p$ is a relatively large number (say, $p=500$) the algorithm runs in slightly sublinear time, and uses logarithmic space. On the other hand, when $p=n^{\varepsilon}$ the algorithm runs in linear time and uses $O(n^{\varepsilon})$ space. For comparison purposes, the usual stack runs in linear time and uses linear space, so the latter case all it consumes more memory without reducing the runtime (from a theoretical point of view).

%

\section{Implementation}\label{sec:implementation}

In this section we introduce our implementation of the stack algorithm framework. 
This section also aims at providing prospective users with practical information that enables them to add the compressed stack to their application or implement their own stack algorithms using the templates provided.

This library was implemented following the C++11 standard, and as such, requires a compiler that can support C++11\footnote{That is one that accepts either -std=c++11 or -std=c++0x compilation flags} \citep{iso2012iec}. As a convention, all the class members (fields or methods) starts with a `\texttt{m}' followed by an upper case and all variables by a lower case. The entirety of pointers used within the library are instantiation of  \texttt{shared\_ptr}\footnote{A smart pointer default class of the C++11 standard that automatically manages the memory of pointers sharing a common element. It follows the \emph{resource acquisition is initialization} (RAII) programming idiom}. The code is available at \citep{baffierCPPcompressed} as an open source library under the MIT license.
A preliminary but operational version --- using Julia language \citep{bezanson2014julia} --- was presented by \citet{baffier2016implementation} and is available as \citep{baffierJULIAcompressed} under the same MIT license.


The rest of this section is organized as follows: First, we describe the file organization of our library in \cref{subsec:classes}. The reading, treatment, and storage of the input data along with the description of the context structure follows in \cref{subsec:types}.
In \cref{subsec:csimpl} we present the compressed stack data structure to be used within a stack algorithm. The whole library is then structured around the stack algorithm class as detailed in \cref{subsec:stackalgo}. Additional functionalities are described in \cref{subsec:extras}. Finally, in \cref{subsec:examples}, we provide a couple of examples instantiating the stack algorithm class.

\subsection{Class and file organization}\label{subsec:classes}

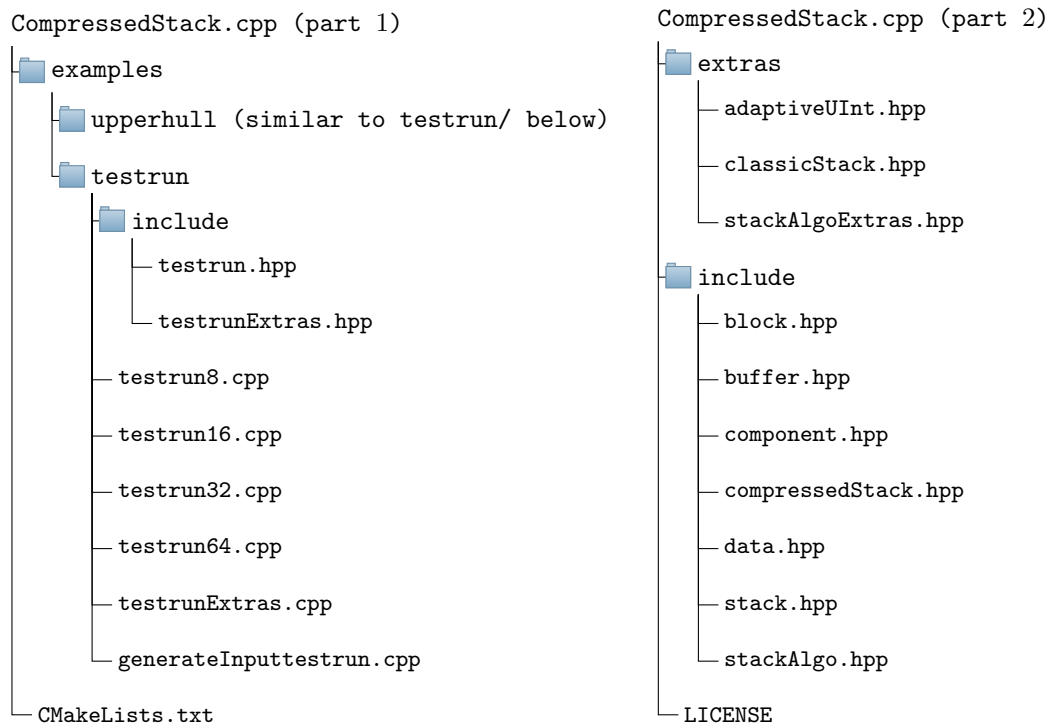
\begin{figure}[ht]
 \begin{forest}
  for tree={
   font=\ttfamily,
   grow'=0,
   child anchor=west,
   parent anchor=south,
   anchor=west,
   calign=first,
   inner xsep=7pt,
   edge path={
    \noexpand\path [draw, \forestoption{edge}]
    (!u.south west) +(7.5pt,0) |- (.child anchor) pic {folder} \forestoption{edge label};
   },
   file/.style={edge path={\noexpand\path [draw, \forestoption{edge}]
    (!u.south west) +(7.5pt,0) |- (.child anchor) \forestoption{edge label};},
    inner xsep=2pt,font=\small\ttfamily
   },
   before typesetting nodes={
    if n=1
    {insert before={[,phantom]}}
    {}
   },
   fit=band,
   before computing xy={l=15pt},
  }
  [CompressedStack.cpp (part $1$)
  [examples
  [upperhull (similar to testrun/ below)]
  [testrun
  [include
  [testrun.hpp,file]
  [testrunExtras.hpp,file]
  ]
  [testrun8.cpp,file]
  [testrun16.cpp,file]
  [testrun32.cpp,file]
  [testrun64.cpp,file]
  [testrunExtras.cpp,file]
  [generateInputtestrun.cpp,file]
  ]
  ]
  [CMakeLists.txt,file]
  ]
 \end{forest}
 \begin{forest}
  for tree={
   font=\ttfamily,
   grow'=0,
   child anchor=west,
   parent anchor=south,
   anchor=west,
   calign=first,
   inner xsep=7pt,
   edge path={
    \noexpand\path [draw, \forestoption{edge}]
    (!u.south west) +(7.5pt,0) |- (.child anchor) pic {folder} \forestoption{edge label};
   },
   file/.style={edge path={\noexpand\path [draw, \forestoption{edge}]
    (!u.south west) +(7.5pt,0) |- (.child anchor) \forestoption{edge label};},
    inner xsep=2pt,font=\small\ttfamily
   },
   before typesetting nodes={
    if n=1
    {insert before={[,phantom]}}
    {}
   },
   fit=band,
   before computing xy={l=15pt},
  }
  [CompressedStack.cpp (part $2$)
  [extras
  [adaptiveUInt.hpp,file]
  [classicStack.hpp,file]
  [stackAlgoExtras.hpp,file]
  ]
  [include
  [block.hpp,file]
  [buffer.hpp,file]
  [component.hpp,file]
  [compressedStack.hpp,file]
  [data.hpp,file]
  [stack.hpp,file]
  [stackAlgo.hpp,file]
  ]
  [LICENSE,file]
  ]
 \end{forest}
 \caption{Directory tree of the CompressedStack.cpp library.}\label{fig:dirtree}
\end{figure}

Our stack algorithm library consists of three main classes: the \texttt{Data} class that handles the input, the \texttt{CompressedStack} class that instantiates transparently a space-optimized stack structure, and the \texttt{StackAlgorithm} class itself --- respectively in \cref{subsec:types,subsec:csimpl,subsec:stackalgo}. These key classes, along with those they depend on, are all grouped in the \texttt{/include} folder as shown in \cref{fig:dirtree}.


Extra functionalities --- including an implementation of a classic stack and some measurement functions --- are in a separate folder (named {\tt /extras}). These classes are not needed for the execution of stack algorithms, but are useful for debugging and evaluation purposes. Finally, examples on how to implement stack algorithms (using normal or compressed stacks) can be found in the {\tt /examples} folder. 

\subsection{Data, context and index types}\label{subsec:types}
In principle, the compressed stack algorithm can work with any kind of input data (depending on the application it could be numbers, points in the plane or edges of a large graph). In order to maintain the versatility, our implementation depends of several parameters provided by the users.

Because we use these abstract types, we cannot precompile the library. The abstract type is implemented in C++ via templates, causing the library to be header-only. The first set of parameters users have to fix relates to the type of input to be read (see \cref{fig:diagramstackinterface}). Specifically, the data type \texttt{D} needs to be provided so the algorithm is aware of the type of elements that it will receive from the input in the main loop of the stack algorithm.

Additionally, the user must fix the \emph{index}. This is a type simply used to fix the maximum expected size of the input (so as to use a reasonable number of bits per input object). Samples for data index up to $8$, $16$, $32$, and $64$ bits unsigned integers are provided in the examples. For the sake of conciseness, the size of the index type is ignored for the rest of this article.

Finally, the  user must fix the \texttt{context}. This context is a data structure of bounded size that is used during the execution of the stack algorithm. The space bottleneck of stack algorithms is the stack itself, but often they use a small amount of memory in their computations (for example, in the convex hull example, context is a bit used to prevent a degenerate case in which the convex hull spirals into itself). This context can be accessed (and modified) at any moment during the execution of the algorithm.


\tikzset{relstyle/.style  =
 {
  font = \footnotesize
}}
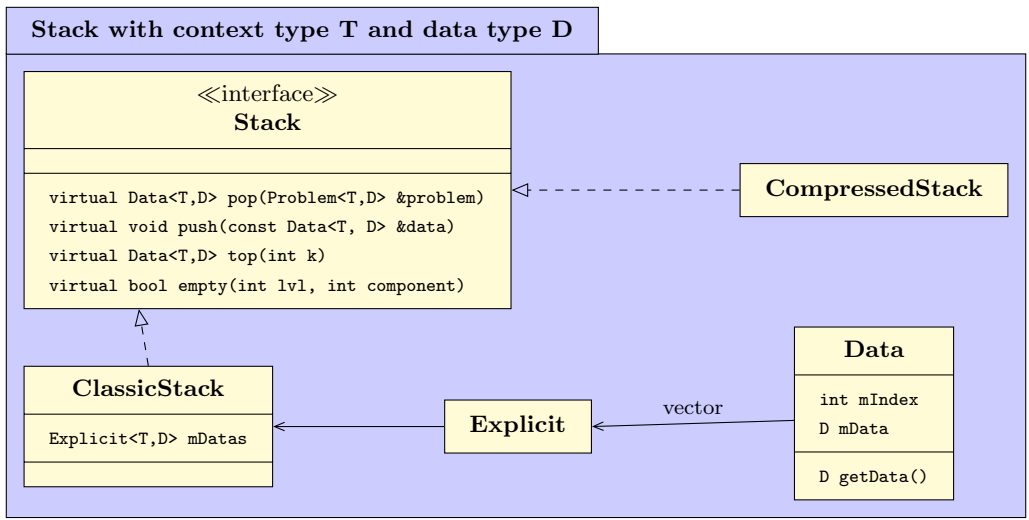
\begin{figure}[t]
 \centering
 \begin{tikzpicture}[scale=0.8]
  \begin{umlpackage}{Stack with context type T and data type D}
   \umlclass[type=interface]{Stack}{}{
    \ts{virtual Data<T,D> pop(Problem<T,D> \&problem)}\\
    \ts{virtual void push(const Data<T, D> \&data)}\\
    \ts{virtual Data<T,D> top(int k)}\\
    \ts{virtual bool empty(int lvl, int component)}
   }
   \umlsimpleclass[right=3cm of Stack.east, anchor=west]{CompressedStack}
   \umlclass[below=.75cm of Stack.south west, anchor=north west]{ClassicStack}{
    \ts{Explicit<T,D> mDatas}
    }{
   }
   \umlsimpleclass[right=2.25cm of ClassicStack]{Explicit}
   \umlclass[below=1.45cm of CompressedStack]{Data}{
    \ts{int mIndex}\\
    \ts{D mData}
    }{
    \ts{D getData()}
   }

   \umluniassoc[relstyle, arg=vector, above, pos=0.5, anchors=-175 and east]{Data}{Explicit}
   \umluniassoc[]{Explicit}{ClassicStack}
   \umlimpl[relstyle, anchors=north and -137]{ClassicStack}{Stack}
   \umlimpl[relstyle, geometry=-|-]{CompressedStack}{Stack}


  \end{umlpackage}
 \end{tikzpicture}
 \caption{Class Diagram for the \texttt{Stack} interface and its implementation by \texttt{ClassicStack}. The \texttt{CompressedStack} implementation is described in \cref{fig:diagramcompressedstack}. The namespace \texttt{std} is used implicitly for \texttt{vector}.}\label{fig:diagramstackinterface}
\end{figure}

\subsection{Compressed Stack}\label{subsec:csimpl}

\begin{figure}[ht]
 \tikzset{Block/.style =
  {
   minimum size=.5cm,
   draw,
   color = BrickRed,
   thick
  }
 }
 \tikzset{Explicit/.style =
  {
   draw,
   inner sep=2mm,
   color = DarkBlue,
   thick
  }
 }
 \tikzset{Level/.style =
  {
   draw,
   inner sep=2mm,
   color = OliveGreen,
   thick
  }
 }
 \tikzset{Component/.style =
  {
   draw,
   inner sep=2mm,
   color = Gray,
   ultra thick
  }
 }
 \tikzset{Edge/.style =
  {
   ->,
   dashed,
   thick
  }
 }
 \tikzset{ToBlock/.style =
  {
   ->,
   thick
  }
 }

 \centering
 \begin{tikzpicture}[scale=1]

  \node (DC1L1) {\ldots};
  \node[above = -0.2cm of DC1L1] {\scriptsize $p$ blocks};
  \node[left = 0.35cm of DC1L1, Block] (FC1L1) {};
  \node[right = 0.35cm of DC1L1, Block] (LC1L1) {};
  \node[Level,fit=(FC1L1) (LC1L1)] (C1L1) {};

  \node[below = 0.35cm of DC1L1] (C1E1i) {\vdots};

  \node[below = 0.5cm of C1E1i] (DC1Li) {};
  \node[left = 0cm of DC1Li, Block] (SC1Li) {};
  \node[right = 0.1cm of SC1Li] {\ldots};
  \node[left = 0.6cm of DC1Li, Block] (FC1Li) {};
  \node[right = 0.6cm of DC1Li, Block] (LC1Li) {};
  \node[Level,fit=(FC1Li) (LC1Li)] (C1Li) {};

  \node[below = 1.25cm of C1Li] (DC1Lj) {\ldots};
  \node[left = 0.35cm of DC1Lj, Block] (FC1Lj) {};
  \node[right = 0.35cm of DC1Lj, Block] (LC1Lj) {};
  \node[Level,fit=(FC1Lj) (LC1Lj)] (C1Lj) {};

  \node[below = 0.35cm of DC1Lj] (C1Ejk) {\vdots};

  \node[below = 0.5cm of C1Ejk] (DC1Lk) {};
  \node[left = 0cm of DC1Lk, Block] (SC1Lk) {};
  \node[right = 0.1cm of SC1Lk] {\ldots};
  \node[left = 0.6cm of DC1Lk, Block] (FC1Lk) {};
  \node[right = 0.6cm of DC1Lk, Block] (LC1Lk) {};
  \node[Level,fit=(FC1Lk) (LC1Lk)] (C1Lk) {};

  \node[below = 1.1cm of C1Lk] (DC1Lh) {\scriptsize ~\{$p$ explicit values\}~};
  \node[Explicit,fit=(DC1Lh)] (C1Lh) {};

  \node[Component,label={above,align=center}:Component 1\\(\texttt{mFirst}),fit=(C1L1) (C1Lh)] (C1) {};

  \node[right = 3cm of DC1L1] (DC2L1) {\ldots};
  \node[left = 0.35cm of DC2L1, Block] (FC2L1) {};
  \node[right = 0.35cm of DC2L1, Block] (LC2L1) {};
  \node[Level,fit=(FC2L1) (LC2L1)] (C2L1) {};

  \node[below = 0.35cm of DC2L1] (C2E1i) {\vdots};

  \node[below = 0.5cm of C2E1i] (DC2Li) {\ldots};
  \node[left = 0.35cm of DC2Li, Block] (FC2Li) {};
  \node[right = 0.35cm of DC2Li, Block] (LC2Li) {};
  \node[Level,fit=(FC2Li) (LC2Li)] (C2Li) {};

  \node[below = 1.25cm of C2Li] (DC2Lj) {\ldots};
  \node[left = 0.35cm of DC2Lj, Block] (FC2Lj) {};
  \node[right = 0.35cm of DC2Lj, Block] (LC2Lj) {};
  \node[Level,fit=(FC2Lj) (LC2Lj)] (C2Lj) {};

  \node[below = 0.35cm of DC2Lj] (C2Ejk) {\vdots};

  \node[below = 0.5cm of C2Ejk] (DC2Lk) {\ldots};
  \node[left = 0.35cm of DC2Lk, Block] (FC2Lk) {};
  \node[right = 0.35cm of DC2Lk, Block] (LC2Lk) {};
  \node[Level,fit=(FC2Lk) (LC2Lk)] (C2Lk) {};

  \node[below = 1.1cm of C2Lk] (DC2Lh) {\scriptsize ~\{$p$ explicit values\}~};
  \node[Explicit,fit=(DC2Lh)] (C2Lh) {};

  \node[Component,label={above,align=center}:Component 2\\(\texttt{mSecond}),fit=(C2L1) (C2Lh)] (C2) {};

  \node[left = 3.5cm of DC1L1, anchor = west] (L1) {Level $1$};
  \node[left = 3.75cm of DC1Li, anchor = west] (Li) {Level $i$};
  \node[left = 3.5cm of DC1Lj, anchor = west] (Lj) {Level $i+1$};
  \node[left = 3.75cm of DC1Lk, anchor = west] (Lk) {Level $h-1$};
  \node[left = 2.65cm of DC1Lh, anchor = west] (Lh) {Level $h$};
  \node[below = 0cm of L1.south west, anchor = north west, align = left] (SL1) {\scriptsize $\frac{n}{p}$ p.c.\ elements};
  \node[below = 0cm of Li.south west, anchor = north west, align = left] (SLi) {\scriptsize $p^{h-i+1}$ \\\scriptsize p.c. elements};
  \node[below = 0cm of Lj.south west, anchor = north west, align = left] (SLj) {\scriptsize $p^{h-i}$ \\\scriptsize p.c. elements};
  \node[below = 0cm of Lk.south west, anchor = north west, align = left] (SLk) {\scriptsize $p^{2}$ \scriptsize p.c. elements};

  \node[right = 1cm of C2L1.north east, minimum size=.5cm, anchor = north west] (FT) {};
  \node[right = 2.5cm of FT, minimum size=.5cm] (LT) {};
  \node[Block,inner sep=2mm,label={above,align=center}:Fully Compressed Tail\\(\texttt{mCompressed}),fit=(FT) (LT)] (T) {\vspace{.025cm}\\ \small\textcolor{black}{$n\cdot\frac{p-2}{p}\approx n$ elements}};

  \node[below = 4.8cm of FT.west, anchor = west, xshift = .5cm] (F) {\scriptsize $:>$ first index};
  \node[below = 0.4cm of F.west, anchor = west] (L) {\scriptsize $:>$ last index};
  \node[below = 0.4cm of L.west, anchor = west] (S) {\scriptsize $:>$ stream position};
  \node[below = 0.4cm of S.west, anchor = west] (C) {\scriptsize $:>$ context};
  \node[below = 0.4cm of C.west, anchor = west] (B) {\scriptsize $:>$ buffer};
  \node[Block, inner sep=2mm,label={above,align=center}:Generic Block,fit=(F) (S) (C) (B)] (GB) {};

  \draw[Edge, bend right] (GB.40) to (T.south);
  \draw[Edge] (GB.180) to (LC2Lj.south east);

  \draw[ToBlock] (C1Lj) to (FC1Li.south);
  \draw[ToBlock] (C2Lj.north) to (SC1Li.south);
  \draw[ToBlock] (C1Lh) to (FC1Lk.south);
  \draw[ToBlock] (C2Lh.north) to (SC1Lk.south);

 \end{tikzpicture}
 \caption{General sketch of a Compressed Stack: red boxes are blocks, green boxes are levels (vector of blocks), blue boxes are explicit values, plain arrows shows the partial compression (p.c.) of a level $i+1$ into a block at level $i$. Recall that $p$ is a parameter set by the user (denoting how much to compress the data), and that $h\approx log_p n$ will denote in how many levels we subdivide the input.}\label{fig:sketchCP}
\end{figure}
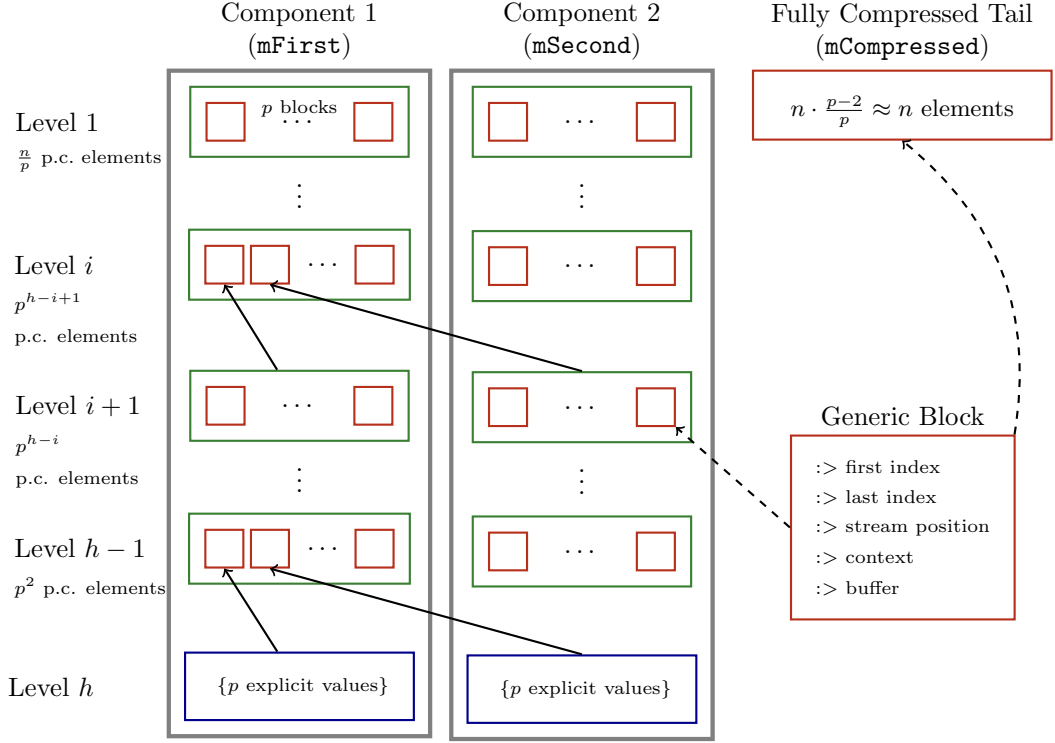

Our implementation pays special attention to modularity, so the stack algorithm is \emph{transparent} to the kind of stack that is actually being used. That is, the algorithm sends push and pop requests and need not know if they are being handled by a regular or a compressed stack. All modifications needed to deal with space constraints are handled by the compressed stack class.

A stack, as shown by the \texttt{Stack} interface in \cref{fig:diagramstackinterface}, must provide a pop and a push function. It might also provide functions to check the top $k$ element of the stack (for some small $k>0$ given by the user). For convenience, a function to test the emptiness of a stack is also expected.
Both the \texttt{CompressedStack} and the \texttt{ClassicStack} classes provided in this library instantiate this \texttt{Stack} interface.

Please see \cref{fig:sketchCP} for an overview of the working of the CompressedStack class. The key trick of saving space is to partition the input into blocks, that are recursively partitioned into blocks, and so on. Recall that a block may be stored \emph{explicitly} (if it stores all elements of the block that have been pushed into the stack), \emph{partially compressed} (in this case, we store information required to reconstruct portions of the block) or \emph{fully compressed} (we store information required to reconstruct the block fully). This data structure compresses information that is unlikely to be accessed in the near future. Depending on the value of parameter $p$ (set by the user) we may compress more or fewer blocks.


\tikzset{relstyle/.style  =
	{
		font = \footnotesize
}}
\begin{figure}[t]
	\centering
	\begin{tikzpicture}[scale=0.8]
		\begin{umlpackage}{CompressedStack with context type T and data type D}
			\umlclass[]{CompressedStack}{
				\ts{int mSize}\\
				\ts{int mSpace}\\
				\ts{int mDepth}\\
				\ts{streampos mPosition}\\
				\ts{shared\_ptr<T> mContext}\\
				\ts{Component<T,D> mFirst}\\
				\ts{Component<T,D> mSecond}\\
				\ts{Level<T,D> mCompressed}\\
				\ts{Buffer<T,D> mBuffer}
				}{
			}
			\umlsimpleclass[below=1cm of CompressedStack.south west, anchor=north west]{Level}
			\umlsimpleclass[right=1.5cm of Level]{Levels}
			\umlclass[below=1cm of Level.south west, anchor=north west]{Block}{
				\ts{int mFirst}\\
				\ts{int mLast}\\
				\ts{streampos mPosition}\\
				\ts{shared\_ptr<T> mContext}\\
				\ts{Buffer<T,D> mBuffer}
				}{
			}

			\umlclass[right=3cm of CompressedStack.north east, anchor=north west]{Component}{
				\ts{Levels<T,D> mPartial}\\
				\ts{ExplicitPointer<T,D> mExplicit}\\
				\ts{Block<T,D> mSign}
				}{
			}
			\umlclass[right=1.75cm of Levels.south east, anchor=south west]{Buffer}{
				\ts{int mSize}\\
				\ts{int mStart}\\
				\ts{ExplicitPointer<T,D> mExplicit}
				}{
				\ts{void pop(SPData<T,D> elt)}\\
				\ts{void push(SPData<T,D> elt)}
			}
			\umlclass[right=1.6cm of Block.south east, anchor=south west]{Data}{
				\ts{int mIndex}\\
				\ts{D mData}
				}{
				\ts{D getData()}
			}
			\umlsimpleclass[right=3cm of Data]{ExplicitPointer}

			\umluniassoc[relstyle, arg = vector$<$shared\_ptr$>$, pos = 0.5]{Data}{ExplicitPointer}
			\umluniassoc[relstyle, arg = vector, pos = 0.5, anchors=122 and south]{Block}{Level}
			\umluniassoc[relstyle, arg = vector, pos = 0.45]{Level}{Levels}
			\umluniassoc[geometry=|-, arm1=3]{ExplicitPointer}{Component}
			\umluniassoc[geometry = |-|]{ExplicitPointer}{Buffer}
			\umluniassoc[relstyle, anchors=north and -114]{Level}{CompressedStack}
			\umluniassoc[relstyle, geometry = |-, anchors=50 and -160]{Levels}{Component}
			\umluniassoc[relstyle, geometry = -|-, anchors = west and -30]{Buffer}{CompressedStack}
			\umluniassoc[relstyle, anchors=west and 30.975]{Component}{CompressedStack}
			\umluniassoc[relstyle, geometry = |-, anchors =-130 and 35]{Buffer}{Block}

		\end{umlpackage}

	\end{tikzpicture}
	\caption{Class Diagram for the CompressedStack and related classes. The namespace \texttt{std} is used implicitly for vector, shared\_ptr, and streampos.}\label{fig:diagramcompressedstack}
\end{figure}
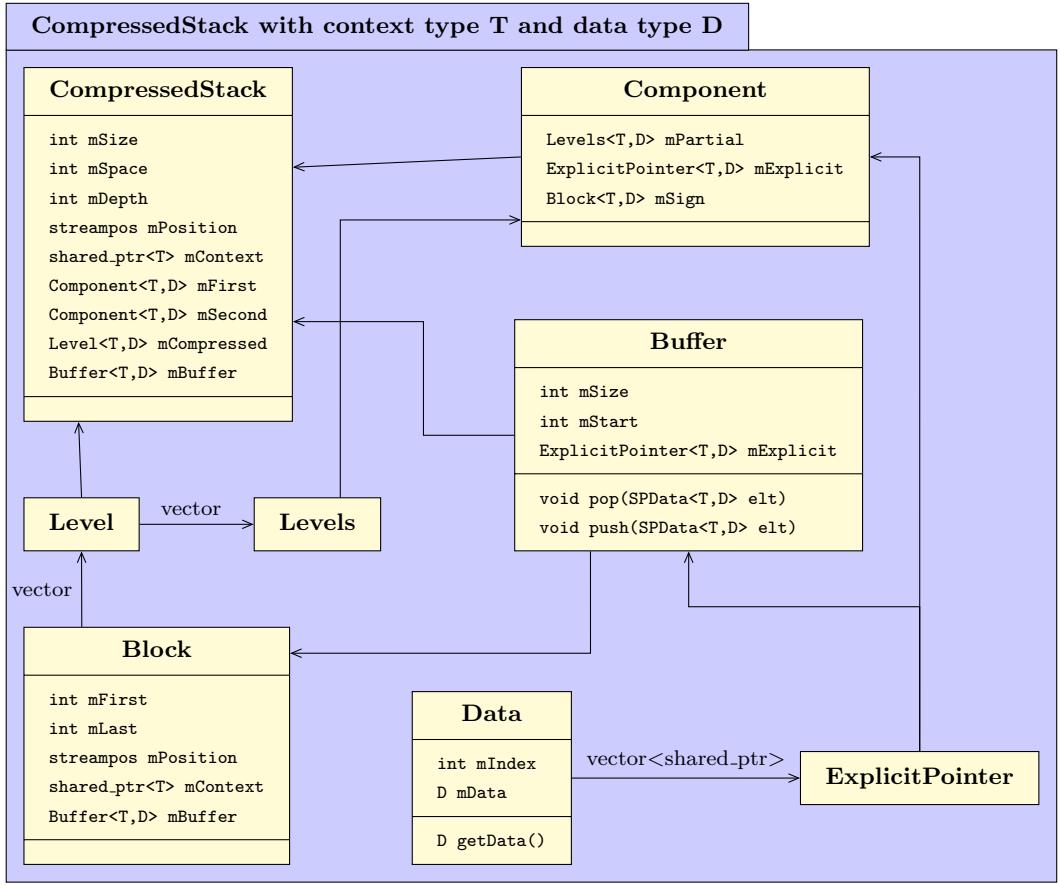

The resulting object-programming structure that constitutes the \texttt{CompressedStack} class is described in \cref{fig:diagramcompressedstack}. The components store compressed data in their \texttt{mPartial} attribute. This takes the form of a matrix (vector of vectors) of blocks. Explicit data is stored in the \texttt{mExplicit} attribute. A \emph{buffer} \texttt{mBuffer} stores the $k$ top elements of a compressed stack that are required to be accessible by the user.

An object of the \texttt{CompressedStack} class includes the following attributes: The number of elements\footnote{The case where the exact number is unknown is an additional functionality that is treated and explained in \cref{subsec:extras}. Although there might be some efficiency loss, the behaviour of the compressed stack is unchanged.} in the input $\I$ as \texttt{mSize}. The space order $p$ as \texttt{mSpace} and the depth $h$ of each components as \texttt{mDepth}.
A pointer to the current position in the stream, \texttt{mPosition} and a pointer to the current context \texttt{mContext}.
The two necessary components of the stack that stores both compressed and explicit data: \texttt{mFirst} and \texttt{mSecond}.
The fully compressed tail of elements \texttt{mCompressed}.

As described in \cref{sec:preliminaries}, a compressed stack may need to self-reconstruct part of its compressed content that is stored in a \emph{block}. This object is implemented as the \texttt{Block} class that stores the index of the first and last elements of the block. It also stores the position in the input stream, the context and an optional buffer.

\subsection{Stack algorithm: Practical vs.\ Theoretical}\label{subsec:stackalgo}
In addition to the compressed stack data structure, we provide a framework for the compressed stack algorithms themselves (i.e., the general scheme described in \cref{algo:theory}). Although their theoretical framework is complete and sound, we introduce some minor variations that may help in practical applications. See the modified version in \cref{algo:practical}.

\begin{algorithm}[!htp]
 \caption{Implementation of a stack algorithm}\label{algo:practical}
 \SetKwInOut{Input}{input}

 \Input{A stack algorithm $\A$ with: an empty stack $\S$; an empty context $\C$; an input stream $\I$.}

 $\A$.\init()\nllabel{line:init}\\
 \While{$\I$.\eof() == false\nllabel{line:startstep}}{
  $a \leftarrow $ $\I$.\readin()\nllabel{line:read}\\
  \While{$\S \neq \emptyset$\nllabel{line:startpop}}{
   \eIf{$\A$.\popc($a$)}{
    $\A$.\prepop(a)\\
    $a' \leftarrow$ $\S$.\pop()\\
    $\A$.\postpop($a$, $a'$)
    }{
    $\A$.\nopop($a$)\\
    \texttt{break} \quad\quad// \emph{exit the while loop}
   }
   }\nllabel{line:endpop}
  \eIf{$\A$.\pushc($a$)\nllabel{line:startpush}}{
   $\A$.\prepush($a$)\\
   $\S$.\push($a$)\\
   $\A$.\postpush($a$)
   }{
   $\A$.\nopush($a$)
   }\nllabel{line:endpush}
  }\nllabel{line:endstep}
 $\A$.\report()\nllabel{line:report}
\end{algorithm}

Consider a stack algorithm $\A$ with associated stack $\S$ (classic or compressed) and context $\C$. First, we describe the key features common to both \cref{algo:theory,algo:practical}:
\begin{itemize}
 \item \init: initialize, if necessary, the empty stack $\S$ and context $\C$. If the algorithm is allowed access to the top $k$ elements of $\S$, then the execution of this function must provide at least $k$ elements of $\I$ into $\S$.
 \item \popc: returns \texttt{true} if the top of the stack $\S$ has to be popped.
 \item \pop: pops the top element of $\S$.
 \item \pushc: returns {\tt true} if the last read element of $\I$ should be pushed into the stack.
 \item \push: pushes that element into $\S$.
 \item \report: $\A$ execute a set of actions aimed towards reporting the objective of the algorithm once all the input data of $\I$ has been processed. In most cases it explicitly returns the contents of the stack (report the top, pop from the stack, report the next one, and so on until the stack is empty).
\end{itemize}

Observe that the scheme of \cref{algo:theory} hides practical aspects, such as how to access the input values $\I$ (on
\cref{line:theoryread}), or additional actions that could be executed together with either the push or pop (\cref{line:theorypop,line:theorypush}).
We provide these additional operations in \cref{algo:practical} defined as follows:
\begin{itemize}
 \item \readin: operation executed each time an input element is read $\I$ (e.g.\ if the input is provided in the form of an input file, this operation processes one line from the file and transforms it into data type $D$).
 \item \prepush{} and \prepop: set of actions to be executed before doing a push and pop, respectively.
 \item \postpush{} and \postpop: set of actions to be executed after a push and pop, respectively.
 \item \nopush{} and \nopop: set of actions to be executed when there is no push or pop, respectively.
\end{itemize}

The user can tune all the functions described above as needed. Other than \readin, by default the procedures do nothing (and can remain as such if not needed by the stack algorithm).
Finally, the algorithms uses an End-Of-File operation \eof{}. This operation is used in \cref{algo:practical} at \cref{line:startstep} and simply returns \texttt{true} once the end of input $\I$ is reached (\texttt{false} otherwise).

All these functions and all basic stack functionalities are summarized in \cref{fig:diagramstackalgo}.

\tikzset{relstyle/.style  =
 {
  font = \footnotesize
}}
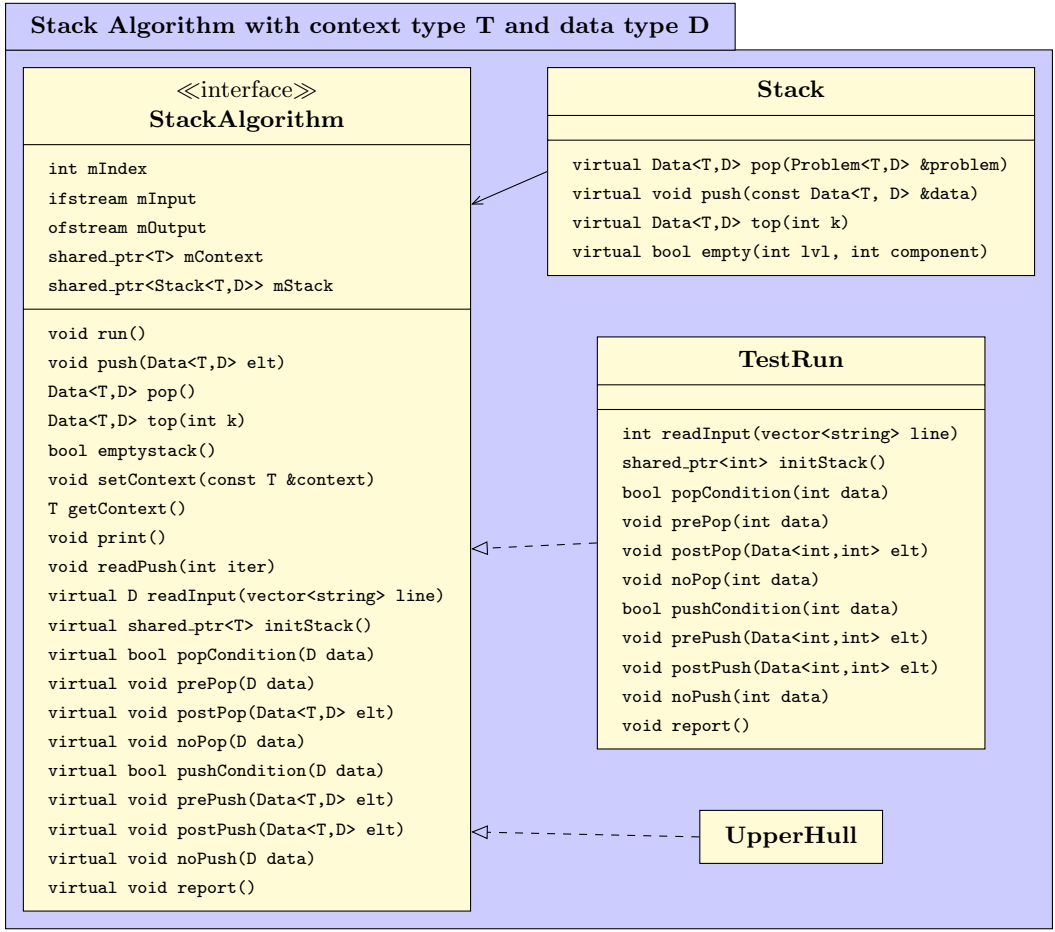
\begin{figure}[t]
 \centering
 \begin{tikzpicture}[scale=0.8]
  \begin{umlpackage}{Stack Algorithm with context type T and data type D}
   \umlclass[type=interface]{StackAlgorithm}{
    \ts{int mIndex}\\
    \ts{ifstream mInput}\\
    \ts{ofstream mOutput}\\
    \ts{shared\_ptr<T> mContext}\\
    \ts{shared\_ptr<Stack<T,D>> mStack}
    }{
    \ts{void run()}\\
    \ts{void push(Data<T,D> elt)}\\
    \ts{Data<T,D> pop()}\\
    \ts{Data<T,D> top(int k)}\\
    \ts{bool emptystack()}\\
    \ts{void setContext(const T \&context)}\\
    \ts{T getContext()}\\
    \ts{void print()}\\
    \ts{void readPush(int iter)}\\
    \ts{virtual D readInput(vector<string> line)}\\
    \ts{virtual shared\_ptr<T> initStack()}\\
    \ts{virtual bool popCondition(D data)}\\
    \ts{virtual void prePop(D data)}\\
    \ts{virtual void postPop(Data<T,D> elt)}\\
    \ts{virtual void noPop(D data)}\\
    \ts{virtual bool pushCondition(D data)}\\
    \ts{virtual void prePush(Data<T,D> elt)}\\
    \ts{virtual void postPush(Data<T,D> elt)}\\
    \ts{virtual void noPush(D data)}\\
    \ts{virtual void report()}
   }
   \umlclass[right=1cm of StackAlgorithm.north east, anchor=north west]{Stack}{}{
    \ts{virtual Data<T,D> pop(Problem<T,D> \&problem)}\\
    \ts{virtual void push(const Data<T, D> \&data)}\\
    \ts{virtual Data<T,D> top(int k)}\\
    \ts{virtual bool empty(int lvl, int component)}
   }
   \umlclass[below=.8cm of Stack.south, anchor=north]{TestRun}{}{
    \ts{int readInput(vector<string> line)}\\
    \ts{shared\_ptr<int> initStack()}\\
    \ts{bool popCondition(int data)}\\
    \ts{void prePop(int data)}\\
    \ts{void postPop(Data<int,int> elt)}\\
    \ts{void noPop(int data)}\\
    \ts{bool pushCondition(int data)}\\
    \ts{void prePush(Data<int,int> elt)}\\
    \ts{void postPush(Data<int,int> elt)}\\
    \ts{void noPush(int data)}\\
    \ts{void report()}
   }

   \umlsimpleclass[below=.8cm of TestRun.south, anchor=north]{UpperHull}

   \umluniassoc[relstyle, anchors=west and 52.99]{Stack}{StackAlgorithm}
   \umlimpl[relstyle, anchors=west and -15.75]{TestRun}{StackAlgorithm}
   \umlimpl[relstyle, anchors=west and -57.5]{UpperHull}{StackAlgorithm}
  \end{umlpackage}
 \end{tikzpicture}
 \caption{Class Diagram for the StackAlgorithm interface corresponding to \cref{algo:practical}. The class TestRun is an implementation of StackAlgorithm for \cref{prob:testrun} where the context type $T = int$ and the data type $D = int$. The namespace \texttt{std} is used implicitly for ifstream, ofstream, shared\_ptr, vector, and string.}\label{fig:diagramstackalgo}
\end{figure}

\subsection{Additional Functionalities}\label{subsec:extras}
This section covers some additional functionalities of our library. 
Most of these functions aim to facilitate the evaluation and possible debugging of the library by the users. Documentation is available on \citep{baffierCPPcompressed}.

Directly into the \texttt{StackAlgorithm} class we count the number of reconstructions executed during its execution. In addition, we provide funcionalities to automatically handle an unknown input size (in a way that is transparent to the user). 
Ideally, the user should input a value $n_{expect}$ that is somehow close to $n$. The compressed stack will work correctly regardless of the value given, but the more accurate the estimate is, the better the compressed stack will perform (if $n \ll n_{expect}$, the algorithm will be much slower, whereas $n_{expect} \ll n$ will cause too much memory to be used).

We also provide the option of using a classic stack (by itself or in parallel to the compressed stack). We also added a function to check that both the classic and the compressed stacks behave similarly. The comparison between the behaviour of the two kind of stacks checks, after each \readin, \pop, and \push operation in the stack, that all information explicitly stored in the compressed stack matches the classic one. 

We encourage the reader to add debugging and analysing tools within this extra framework and keep the original compressed stack class as light as possible. Contributions to \citep{baffierCPPcompressed} are welcome.

\subsection{Example of Stack Algorithms}\label{subsec:examples}
To illustrate the use of the compressed stack and of the stack algorithm, we provide two examples. This will also showcase how easy it is to implement new stack algorithms. The first example (\cref{prob:testrun}) is a minimal one that can simulate any distribution of pops and pushes that we use in the experiments of \cref{sec:xp}. The upper hull problem introduced in \cref{subsec:hull_descri} is referred as \cref{prob:convexhull}. Both are fully implemented and available on \citep{baffierCPPcompressed}.
We also provide an instance generator for both problems.

\begin{problem}[Test Run]\label{prob:testrun}
This is an artificial stack algorithm that executes push and pop operations for debugging purposes. The data type \texttt{D} is a pair of positive integers separated by a comma. The first number indicates the value to be pushed into the stack whereas the second indicates the number of pops that should be done in lines~\ref{line:startpop}-\ref{line:endpop} of \cref{algo:practical}. The purpose of this algorithm will be clear in \cref{sec:xp}.
\end{problem}


\begin{figure}[th!]
 \begin{lstlisting}[frame=bt]
  std::shared_ptr<emptyContext> initStack() {
    // first, read and push two values
    StackAlgo<emptyContext, Point2D, int>::readPush(2);
    // then initialize context (which in this case is NULL)
    std::shared_ptr<emptyContext> context;
    return context;
  }

  bool popCondition(Point2D last) {
    Point2D minus1, minus2;
    // read the two previous elements
    minus1 =
      StackAlgo<emptyContext, Point2D, int>::top(1).getData();
    if (StackAlgo<emptyContext, Point2D, int>::
                              mStack->getBufferLength() < 2) {
      return true;
    }
    minus2 = StackAlgo<emptyContext, Point2D, int>::
                                             top(2).getData();
    // Pop condition is true depending on the points position
    if (Point2D::orientation(minus2, minus1, last) == 1) {
      return true;
    }
    return false;
  }

  bool pushCondition(Point2D data) {
    return true;
  }
 \end{lstlisting}
 \caption{Instation of a \texttt{StackAlgorithm} template for the upper hull problem. Although we omit the code of the \texttt{Point2D::orientation} function, this implementation of the upper hull is concise, simple and independant of the kind of stack selected by the user.}\label{code:upperhull}
\end{figure}

\begin{problem}[Upper Hull]\label{prob:convexhull}
This algorithm computes the upper hull of a set of points in the plane, assuming that the input is given in increasing values of the $x$-coordinate. As shown in \cref{subsec:hull_descri}, this can be done with a stack algorithm. For this problem, the data type \texttt{D} is a class \texttt{Point2D} that represents two dimensional points, and the input is simply the list of points as a pair of \texttt{float} coordinates. The output is an ordered list of points which represent the vertices of the upper hull.

The algorithm of Lee does not need any context to compute the upper hull of a set of points that are sorted in the $x$-axis\footnote{If the points come sorted in a different way, then a context will be needed to prevent some degenerate cases. In any case, these degenerate cases have no impact on the usage of the stack, so we ignore them.}, so it is simply set to a null pointer in C++11. Note that, during the execution of the algorithm, it will access the top two elements of the stack when determining whether or not to pop.

Since this algorithm fits in the stack algorithms class, it can be easily implemented (see a concise implementation in \cref{code:upperhull}).
\end{problem}

\section{Experimental Results}\label{sec:xp}

This section contains experimental studies aimed at analysing the practical performance of the compressed stack.

\subsection{Data and evaluation measures.}

In our experiments we use synthetic data. The main reason behind this decision is that it allows us to fully control the conditions in which the compressed stack would have to work (for example, we can control the number of reconstructions that will be executed). In addition, compressed stack algorithms spend a significant amount of time computing things that are unrelated to the stack (for example, in the upper hull algorithm, the \popc\ operation must compute the determinant of a $3\times 3$ matrix). Since these additional computations can affect the leading term for the running time, it make it hard to compare the performance of both stacks.

Our aim is to  measure the differences between using a regular and a compressed stack. Thus, it is desirable to keep additional overhead to a minimum. As such, we implemented our experiments with the testrun problem described as \cref{prob:testrun} in \cref{sec:implementation}.

The first experiment represents a very favourable situation for the compressed stack: a case in which data is accessed almost sequentially (and thus it is ideal for compression purposes). The goal in this case is to show the potential of memory saving that this data structure can achieve. The second test aims at setting a much more challenging scenario: continuous pops are set in a way such that scattered positions of the input file need to be accessed. This forces the compressed stack to repeatedly reconstruct portions of the input, and thus potentially lose a lot of time when compared to the classic stack.

In order to measure the performance of our implementation we focused on two magnitudes: the maximum amount of memory used by the algorithm and the running time. To measure the first, we used a heap profiler called {\em massif} \citep{nethercote2006valgrind} belonging to the Valgrind software suite. This software keeps track of the memory allocated in the execution heap at intervals of predefined length and outputs detailed heap memory usage. While this software allowed us to measure maximum memory usage, its use made the running of the algorithms much slower.

As the second magnitude that we were interested in was run-time, we needed to make two separate runs for every test case. In the first execution, we run massif alongside our code, obtaining memory usage data. In the second execution we run the test code alone in order to obtain unadulterated run-time readings. Consequently, in this section we present memory usage readings as outputted by massif (in bytes unless otherwise stated) as well as the times of the algorithms (in seconds) when run without massif. In order to be able to present values for widely different sizes, in each execution, we doubled the size of the input $n$ (of size $n=2^i$ for increasing values of $i$).

For comparison purposes, we also coded a classical stack class that implements the stack interface described in \cref{fig:diagramstackinterface}. This is simply a wrapper class for the widely used C++ std vector that implements the \texttt{std::stack} interface. We believe that this is a representative instance of a classical stack using unconstrained memory. Regarding the parameters of the compressed stack, we focused on the `$p$' space parameter introduced in \cref{subsec:cpoverview}. For the purposes of this experiment, it suffices to know that the larger $p$ is, the more space is used by the compressed stack (and fewer reconstructions are needed).

In order to illustrate the effect of this parameter in the performance of the compressed stack, we present results for eight different values of $p$. Specifically, the first four values are fixed (10, 50, 100 and 500) while the other four change with the size of the input $n$: $\sqrt{n}$,$\sqrt[4]{n}$,$\sqrt[8]{n}$ and $\log{n}$. Fixed values allow us to illustrate how an imbalance between $n$ and $p$ may result in very high running times for the compressed stack while the varying (and ever growing) values $\sqrt{n}$,$\sqrt[4]{n}$,$\sqrt[8]{n}$ and $\log{n}$ exemplify the trade-off between a lower memory limit for the compressed stack use and higher computation times. In order to keep the section within reasonable length limits, we only present summary figures of memory usage and running times. Detailed tables can be downloaded at \citep{baffierCPPcompressed}.

\subsection{Linear sized stack}\label{pushonlyEXP}
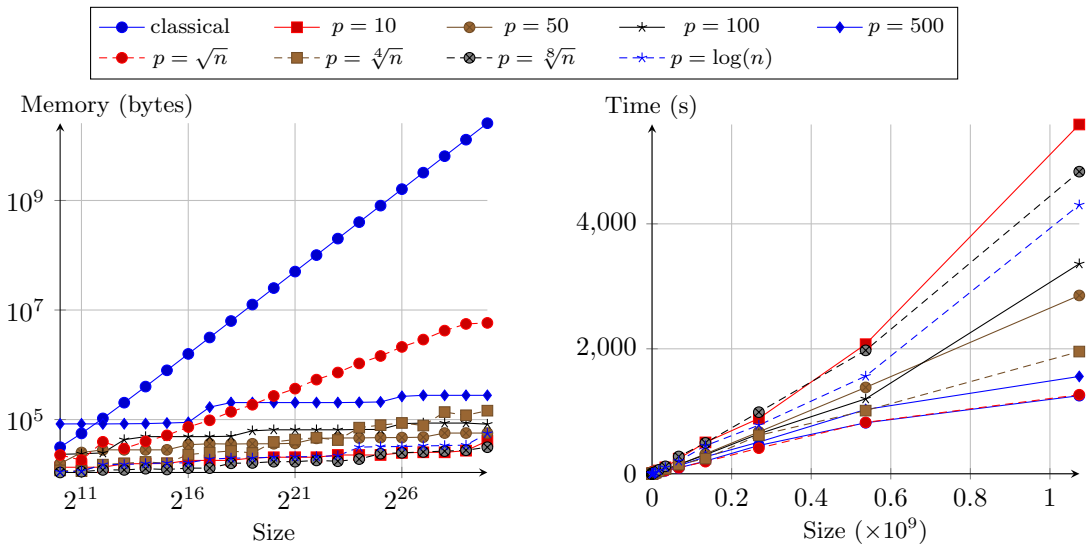
\begin{figure}[t]
 \centering
 \begin{subfigure}[h]{.49\textwidth}
  \centering
  \begin{tikzpicture}[scale=1]
   \begin{axis}[
    width=\textwidth,
    axis x line=bottom,axis y line=left,
    xlabel={\small Size},ylabel={\small Memory (bytes)},
    x label style={at={(axis description cs:0.5,-0.12)}},
    ylabel style={rotate=-90},
    y label style={at={(axis description cs:.335,1.05)}},
    legend entries={\footnotesize classical,\footnotesize $p=10$,\footnotesize $p=50$,\footnotesize $p=100$,\footnotesize $p=500$,\footnotesize $p=\sqrt{n}$,\footnotesize $p=\sqrt[4]{n}$,\footnotesize $p=\sqrt[8]{n}$,\footnotesize $p=\log(n)$},legend style={at={(0.07,1.125)},anchor=south west,legend columns=5,
     /tikz/every even column/.append style={column sep=0.5cm}},
    xmode=log,
    log basis x={2},
    ymode=log,
    log basis y={10},
    grid = major
    ]
    \addplot table [x=size, y=memory, col sep=comma] {data/pushOnlyClassical.csv};
    \addplot table [x=size, y=memory, col sep=comma] {data/pushOnlyCS10.csv};
    \addplot table [x=size, y=memory, col sep=comma] {data/pushOnlyCS50.csv};
    \addplot table [x=size, y=memory, col sep=comma] {data/pushOnlyCS100.csv};
    \addplot table [x=size, y=memory, col sep=comma] {data/pushOnlyCS500.csv};
    \addplot table [x=size, y=memory, col sep=comma] {data/pushOnlyCSsqrt.csv};
    \addplot table [x=size, y=memory, col sep=comma] {data/pushOnlyCS4root.csv};
    \addplot table [x=size, y=memory, col sep=comma] {data/pushOnlyCS8root.csv};
    \addplot table [x=size, y=memory, col sep=comma] {data/pushOnlyCSlog.csv};
   \end{axis}
  \end{tikzpicture}
  \caption{\small Memory comparison classical vs compressed. For ease of visualization, a logarithmic scale was used in both axes.}
  \label{fig:memCompPO}
 \end{subfigure}
 \hfill
 \begin{subfigure}[h]{.49\textwidth}
  \centering
  \vspace{.725cm}
  \begin{tikzpicture}[scale=1]
   \begin{axis}[
    width=\textwidth,
    every x tick scale label/.style={at={(xticklabel cs:2)},anchor=south west},
    every y tick scale label/.style={at={(xticklabel cs:2)},anchor=south west},
    axis x line=bottom,axis y line=left,
    xlabel={\small Size ($\times 10^9$)},ylabel={\small Time (s)},
    x label style={at={(axis description cs:0.5,-0.1)}},
    ylabel style={rotate=-90},
    y label style={at={(axis description cs:.13,1.05)}},
    grid = major
    ]
    \addplot table [x=size, y=time, col sep=comma] {data/pushOnlyClassical.csv};
    \addplot table [x=size, y=time, col sep=comma] {data/pushOnlyCS10.csv};
    \addplot table [x=size, y=time, col sep=comma] {data/pushOnlyCS50.csv};
    \addplot table [x=size, y=time, col sep=comma] {data/pushOnlyCS100.csv};
    \addplot table [x=size, y=time, col sep=comma] {data/pushOnlyCS500.csv};
    \addplot table [x=size, y=time, col sep=comma] {data/pushOnlyCSsqrt.csv};
    \addplot table [x=size, y=time, col sep=comma] {data/pushOnlyCS4root.csv};
    \addplot table [x=size, y=time, col sep=comma] {data/pushOnlyCS8root.csv};
    \addplot table [x=size, y=time, col sep=comma] {data/pushOnlyCSlog.csv};
   \end{axis}
  \end{tikzpicture}
  \caption{\small Time comparison classical vs compressed. No scaling in either axis is done for this figure.}\label{fig:timeCompPO}
 \end{subfigure}
 \caption{As expected from theory, the normal stack uses a linear amount of space whereas the compressed stack only logarithmic. Regarding runtime, the classic stack has, as expected, the best performance of all. For large values of $p$ (such as $p=\sqrt[4]{n}$) the running time is comparable to that of the classic stack. Conversely, for smaller values we can see that running times increase significantly. This is because smaller values of $p$ need more reconstructions, which produce higher computation times.}
 \label{fig:ex_1rf}
\end{figure}
In this first test we aim at creating a scenario that maximised the possibility of memory saving by the compressed stack with minor impact on the runtime. We consider the case in which the stack contains a linear fraction of the input. Specifically, fix a probability $\rho\in [0,1]$; then every element of the input is pushed, and a pop will be executed with probability $1-\rho$. In terms of the testrun problem defined, this stood for an input made up of a text file with a list of pairs of integers. For every pair, the first integer was a random positive number and the second was chosen to be 1 with probability $1-\rho$.

For any fixed $\rho>0$ the expected size of the stack is $\rho n$, and thus the memory used by the classic stack will be linear, whereas the compressed stack will only store a logarithmic amount of elements. \Cref{fig:timeCompPO,fig:memCompPO} show the memory used and runtime of our experiments for the case in which $\rho=1$ (and thus no pops are ever executed). While we acknowledge that this is not a realistic situation, it does highlight the potential saving of space achieved by the compressed stack in cases where a large portion of it does not change.

To simulate more realistic situations, we also repeated the experiment with different values of $\rho$. In all cases, the tendencies observed were similar to the case without pops: the larger the value of $\rho$ the fewer pops are executed, thus the more memory is needed (for example, the compressed stack with $p=\sqrt[4]{n}$ the memory used when $\rho=1$, is between 1.6 and 2 times that of $\rho=0.1$). However, the classic stack always needs a linear amount of space (for example the memory used increases around 24 times for the same parameters). On the other hand, with a larger number of pops, the compressed stack must be reconstructed more often, which brings noticeable increases in the running time (of 1.24 times slower on average on the case mentioned, for example). Since the overall performance for all values of $\rho$ is similar, we refer the interested reader to \citep{baffierCPPcompressed}.

\Cref{fig:memCompPO} depicts the maximum amount of memory needed in this test.  A logarithmic scale (of base 2 for the x axis and base 10 for the y axis) is used for ease of visualization. The figure shows how in this test the classical stack needs much more memory than any of the compressed variants. There are two exceptional cases in which the classical stack uses less memory than a compressed stack algorithm, but it only happens for extremely large values of $p$ when compared to the input size (specifically, the compressed stack with $p=500$ and input sizes of $2^{10},2^{11}$). This shows how the choice of parameter $p$ is important to optimize the performance of the compressed stack. In this case $p$ is too close to $n$ ($n=2^{10}=1024$) and the structure of the stack is wasted as most values are kept explicitly.

The memory needed by the normal stack exceeds that of any compressed variant by two orders of magnitude already by size $2^{19}$. This difference grows together with $n$ (reaching four orders of magnitude for $2^{29}$). The maximum memory needed by the classical stack in the test was over 25 Gigabytes for an input file of size $2^{30}$ (over 1000 million). 
Conversely, the compressed stack, in the worst case, only needed 54.6 Megabytes.

If we compare the different values of $p$ for the compressed stack, we observe that, as expected, smaller values of $p$ result in lower memory usage.
Moreover, the results of algorithms with fixed values of $p$ match the ones of variable $p$ at expected values (for example, the algorithm with $p=500$ should perform like the algorithm with $p=\sqrt{n}$ when $\sqrt{n}=500 \leftarrow n=250000 \approx 2^{19}$, and the two curves meet around that value). For fixed values of $p$ it is easy to see that the memory grows logarithmically (specifically, we see a bump in the memory requirements when $\lceil \log_p n \rceil$ changes), whereas the growth is smoother for variable values of $p$. In all cases, the growth is very monotone, which matches the expectation from theory.

The downside of using a small value of $p$ is that the running time will increase. The effect of the growth of $p$ is only mildly visible in \Cref{fig:timeCompPO,fig:memCompPO} but will be more evident in \cref{CTexp}. The reason for this is that the number of times that the reconstruct function is invoked is small. 

\subsection{A challenging scenario for the compressed stack}\label{CTexp}

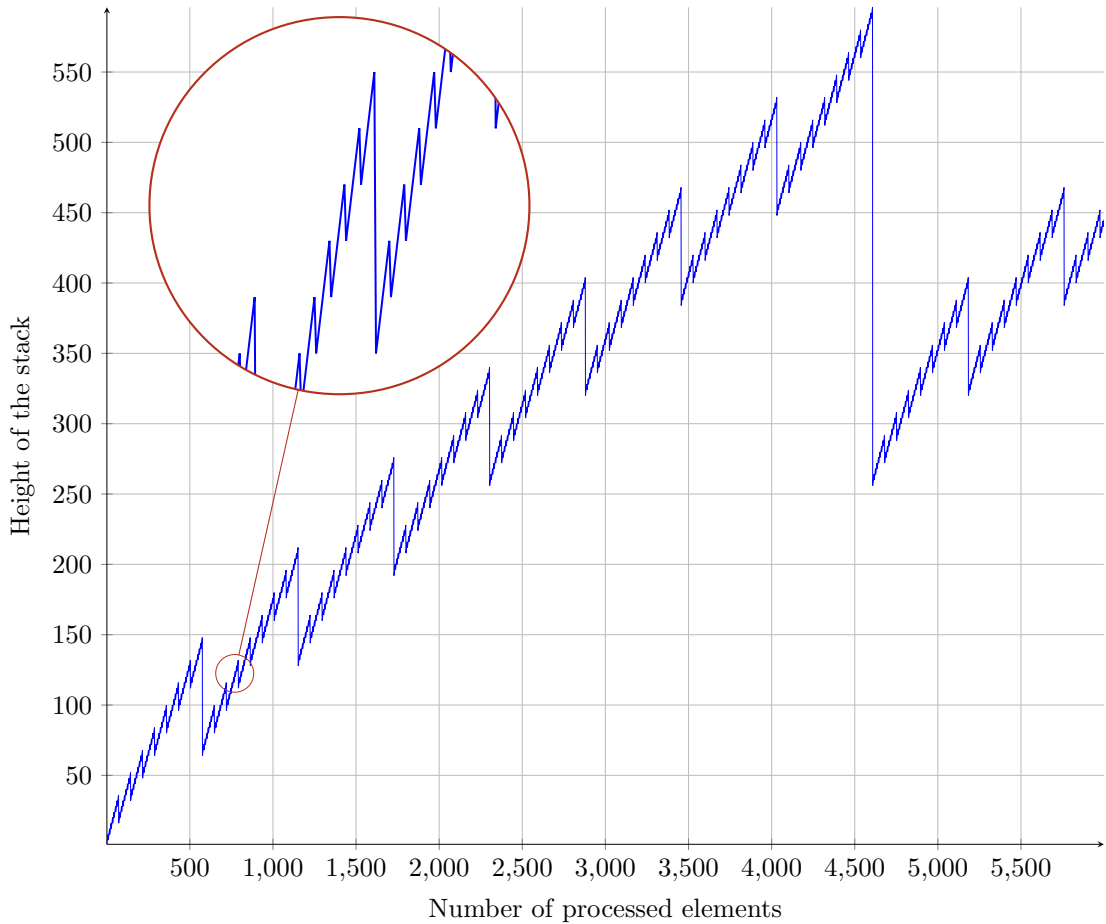
\begin{figure}[tb]
 \centering
 \begin{tikzpicture}[scale=1,spy using outlines= {circle, magnification=10, connect spies}]
  \begin{axis}[
   width=\textwidth,
   axis x line=bottom,axis y line=left,
   xlabel={Number of processed elements},ylabel={Height of the stack},
   restrict x to domain=0:5999,
   grid = major,
   line width = .075pt
   ]
   \addplot table [x="index", y="elements", col sep=comma,mark=none] {data/xmastree.csv};
   \coordinate (spypoint) at (axis cs:770,122.5);
   \coordinate (magnifyglass) at (axis cs:1400,455);

  \end{axis}
  \spy [BrickRed, size=5cm] on (spypoint) in node[fill=white] at (magnifyglass);

 \end{tikzpicture}
 \caption{Number of elements in the compressed stack as a function of the number of processed elements. The zoomed image shows two levels of recursion, and outside the zoom we can see two more levels (for a total of four).}\label{fig:xmastree}
\end{figure}


In this section, we present a different test scenario that specifically tries to maximize the problems of the compressed stack: we set the input to produce push-pop cycles so as to force many reconstructions. Hence, overall the values that are pushed are at non-contiguous positions, making it difficult for the information to be compressed. Moreover, we make sure that the overall data that needs to be stored grows, but not at a linear rate. This is again a very artificial construction, but we believe that it shows that even under difficult conditions the compressed stack performs reasonably well. In order to create this setting, the instance forces the following operations into the stack:

\begin{itemize}
 \item Push 8 elements, pop half of them (4).
 \item Repeat the previous step 8 times. At this point we have processed 64 elements and keep half of them (32) in the stack.
 \item Pop half of the stack, resulting in a stack of $16$ elements.
 \item Repeat this double loop 8 times, resulting in 128 elements in the stack after 512 elements have been processed.
 \item We again pop half of the stack, keeping only 64 elements in the stack.
 \item Repeat now the triple loop 8 times, and so on
\end{itemize}

This procedure keeps adding cycles of increasing length until the desired input size $n$ is reached. The stack stores 4 elements that are consecutive in the input, but the spacing between numbers to store grows exponentially, creating a very difficult situation for the compressed stack (since reconstruction operations will have to scan large portions of the input for just a few elements that are stored explicitly in the regular stack).

As in the previous example, $n$ was taken following the powers of 2. Note that the choice of making loops with 8 iterations (and popping half of the stack at each step) is arbitrary. As 8 is a power of 2, the memory usage patterns in the classical stack become easier to predict (more details below). We call this test the {\em Christmas tree} test (because the height of the stack forms a Christmas tree-like shape, see \cref{fig:xmastree} for a graphical representation on the number of elements present in the stack as a function of the input size.

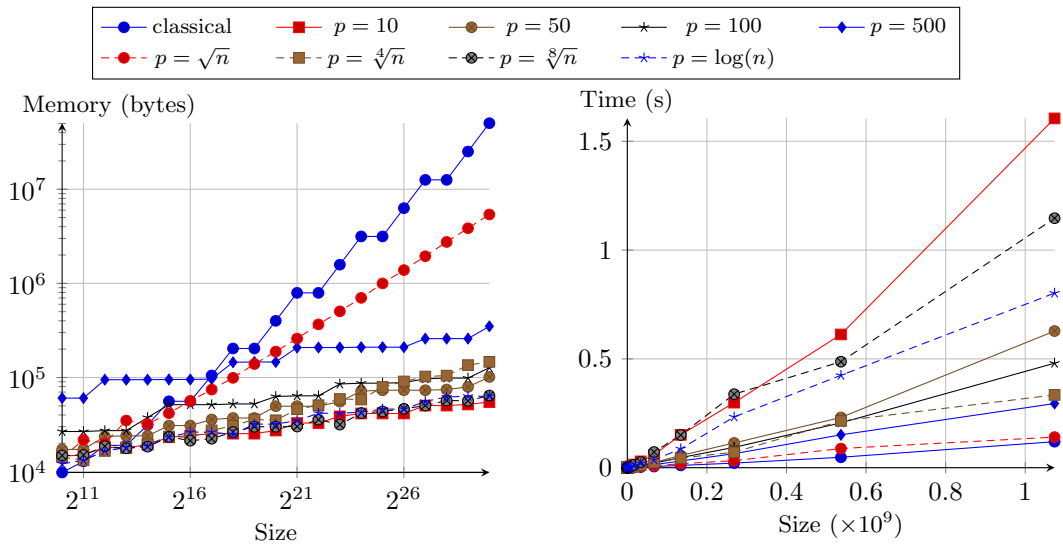
\begin{figure}[t]
 \centering
 \begin{subfigure}[h]{.49\textwidth}
  \centering
  \begin{tikzpicture}[scale=1]
   \begin{axis}[
    width=\textwidth,
    axis x line=bottom,axis y line=left,
    xlabel={\small Size},ylabel={\small Memory (bytes)},
    x label style={at={(axis description cs:0.5,-0.12)}},
    ylabel style={rotate=-90},
    y label style={at={(axis description cs:.335,1.05)}},
    legend entries={\footnotesize classical,\footnotesize $p=10$,\footnotesize $p=50$,\footnotesize $p=100$,\footnotesize $p=500$,\footnotesize $p=\sqrt{n}$,\footnotesize $p=\sqrt[4]{n}$,\footnotesize $p=\sqrt[8]{n}$,\footnotesize $p=\log(n)$},legend style={at={(0.07,1.125)},anchor=south west,legend columns=5,
     /tikz/every even column/.append style={column sep=0.5cm}},
    xmode=log,
    log basis x={2},
    ymode=log,
    log basis y={10},
    grid = major
    ]
    \addplot table [x=size, y=memory, col sep=comma] {data/CTTestNormal.csv};
    \addplot table [x=size, y=memory, col sep=comma] {data/CTTestCS10.csv};
    \addplot table [x=size, y=memory, col sep=comma] {data/CTTestCS50.csv};
    \addplot table [x=size, y=memory, col sep=comma] {data/CTTestCS100.csv};
    \addplot table [x=size, y=memory, col sep=comma] {data/CTTestCS500.csv};
    \addplot table [x=size, y=memory, col sep=comma] {data/CTTestCSsqrt.csv};
    \addplot table [x=size, y=memory, col sep=comma] {data/CTTestCS4root.csv};
    \addplot table [x=size, y=memory, col sep=comma] {data/CTTestCS8root.csv};
    \addplot table [x=size, y=memory, col sep=comma] {data/CTTestCSlog.csv};
   \end{axis}

  \end{tikzpicture}
  \caption{\small Memory comparison classical vs compressed (logarithmic scale used in both axes), CT test.}\label{fig:memCompCT}
 \end{subfigure}
 \hfill
 \begin{subfigure}[h]{.49\textwidth}
  \centering
  \vspace{.95cm}
  \begin{tikzpicture}[scale=1]
   \begin{axis}[
    every x tick scale label/.style={at={(xticklabel cs:2)},anchor=south west},
    every y tick scale label/.style={at={(xticklabel cs:2)},anchor=south west},
    width=\textwidth,
    axis x line=bottom,axis y line=left,
    xlabel={\small Size ($\times 10^9$)},ylabel={\small Time (s)},
    x label style={at={(axis description cs:0.5,-0.1)}},
    ylabel style={rotate=-90},
    y label style={at={(axis description cs:.13,1.05)}},
    grid = major
    ]
    \addplot table [x=size, y=time, col sep=comma] {data/CTTestNormal.csv};
    \addplot table [x=size, y=time, col sep=comma] {data/CTTestCS10.csv};
    \addplot table [x=size, y=time, col sep=comma] {data/CTTestCS50.csv};
    \addplot table [x=size, y=time, col sep=comma] {data/CTTestCS100.csv};
    \addplot table [x=size, y=time, col sep=comma] {data/CTTestCS500.csv};
    \addplot table [x=size, y=time, col sep=comma] {data/CTTestCSsqrt.csv};
    \addplot table [x=size, y=time, col sep=comma] {data/CTTestCS4root.csv};
    \addplot table [x=size, y=time, col sep=comma] {data/CTTestCS8root.csv};
    \addplot table [x=size, y=time, col sep=comma] {data/CTTestCSlog.csv};
   \end{axis}
  \end{tikzpicture}
  \caption{\small Time comparison classical vs compressed, CT test. Linear scale was used.}\label{fig:timeCompCT}
 \end{subfigure}
 \caption{Christmas Tree experiments. In this test, designed to be challenging for the compressed stack, the memory saving respect to the classical stack is much smaller (and in a few instances the classical stack even needs less memory than some compressed stacks). Concerning time, the constant calls to the reconstruct function make the compressed stack much slower (as exemplified, for example by the behavior of the compressed stack with $p=10$. However, even in this tailored scenario, we can see how the compressed stack maintains a capped memory usage as well as relatively low running times if the value of $p$ is chosen appropriately (as can be seen for example, in the values of $p=\sqrt[4]{n}$). }\label{fig:ex_2rf}
\end{figure}

\Cref{fig:memCompCT} shows the memory used in the Christmas tree test by all different stacks. We observe that the memory usage by the classical stack is significantly lower than in the previous experiment. This happens because very few elements are added into the stack (every factor of eight that the input grows, the space requirements only grow by a factor of $4$, hence the stack has roughly $n^{2/3}$ elements).

This memory usage also grows stepwise in the sense that similar memory usages are detected for consecutive sizes. This is caused by the shape of the christmas tree (after a big pop like the one we find at around 4500 in \cref{fig:xmastree} increases in the input size will not increase the amount of memory needed until we reach a bigger loop). 


We observe that the amount of memory needed by the compressed stacks is almost the same as the one needed in the previous experiment (the ratio varies slightly with every instance but stays between 0.8 and 1.2 of each other). The increase of the use of memory by the compressed stack in some cases is produced by the fact that the reconstruct function (frequently invoked in this example) duplicates parts of the input that might be of significant length.

It is important to notice that, although in this case, the memory saved by using the compressed stack is small (and in some of the cases with smaller $n$ the compressed stack even needs more memory than the classical stack), the main reason for this is that the memory needed by the classical stack in this test is low. For example, the highest memory usage value in this test is four orders of magnitude lower that the maximum of the previous test. Even in such an ill conceived example, the maximum memory required by a compressed stack (with $p=\sqrt{n}$) is two orders of magnitude smaller than the classic stack. This shows how the memory used by the compressed stack stays capped even in the worse situations.

As expected, the number of reconstructions is much larger for  the christmas tree instance. As such, the difference in the runtimes between the classic stack and a compressed one grows (as seen in \Cref{fig:timeCompCT}). We observe that, although for small values of $p$ the running times become unfeasible, for larger values the runtime is comparable to the one of a regular stack. For example, for $p=\sqrt[4]{n}$, the runtime in average increases by a factor of 2.32 (with 4.50 in the worst case).

This table exemplifies how the theoretical time-space trade-off is realised in practice and can itself constitute a starting point for prospective users of the compressed stack data structure. The left side of the table would be used to assess how much memory could be needed by every configuration of the compressed stack (expressed in the values of $p$) while the right side would provide an indication of the time penalty that the use of the compressed stack might produce.
For example, in this case we believe the best compromise is obtained by the compressed stack with $p=\sqrt[4]{n}$ although the user should also chose the smaller value of $p$ that fits their memory constraints in order to obtain the fastest running algorithm.

\section{Conclusions}\label{sec:conclusion}

The experiments of this paper show that the compressed stack structure can be very useful not only from a theoretical point of view, but also on a practical level. Parallel to our work, a similar experimental study for another time-space trade-off problem previously only studied from a theoretical standpoint was done by~\citep{wolfgang}. Specifically, they studied the time-space dependency for the the problem of computing the shortest path between two points in a simple polygon. We believe that a trend of similar studies will soon follow for these or related problems.

Other than our preliminary implementation~\citep{baffierJULIAcompressed}, this is the very first implementation of the compressed stack data structure introduced in \citep{barba2014spacetime}. The source code along with the data from the experiments presented in this paper is available for download as \citep{baffierCPPcompressed}. 

The results presented show how the compressed stack, along with other memory constrained algorithms has a huge potential to impact new technological contexts such as sensor networks or mobile phone apps. Specifically, \Cref{pushonlyEXP} presented a (synthetic) situation where a normal stack needed 25 Gigabytes of memory while compressed stack implementations needed at most 55.5 Megabytes. This situation represents a challenge for current desktop computers and is infeasible in even the more advanced mobile phones (Apple's Iphone 7, for example has 2 Gigabytes of RAM memory). Although this was a tailor-made case, it still shows how the property of the compressed stack of being able to limit memory usage by trading it for computation time has the potential of opening new application possibilities.

The reduction of memory of this data structure is undeniable, even in the very unfavourable scenario of the christmas tree. Moreover, the amount of memory needed in all scenarios is very stable (the amount of memory needed between the ideal and unfavorable scenarios lies between 0.9 and 1.2 of each other). This makes the compressed stack very robust at holding memory limitations, even for situations in which we do not know much about the structure of the input. 

Although it was known that the saving of memory implies a runtime penalty, the exact amount was unclear. In this paper we have quantified how big much of a penalty to expect. The experiments show how the behaviour of the compressed stack can be predicted. Special attention has been given to comparing the practical behaviour to theoretical predictions. 
For example, the best value of $p$ from a theoretical point of view is a fixed large constant as it gives the best time-space product. However, in our examples we have seen that a fixed value does not always perform well (for too small instances it may consume even more memory than the classic stack, and for larger instances the runtime may increase too much). Instead, values that depends on the size of the input (such as $\sqrt[4]{n}$) provide the best practical behaviour. 

Each user of the compressed stack can choose the value of $p$ so it fits the memory constraints in each specific situation and have an assessment on what the increase in run time will be.
The examples presented in this paper can be used as guidelines by prospective users of the data structure to chose the values of $p$ when they implement their applications using the stack algorithm templates provided in our library.

\bibliographystyle{abbrvnatdoi}
\bibliography{bibliography}

\end{document}